\documentclass[aip,jap,preprint, amsmath,amsfonts,superscriptaddress]{revtex4-1} 
\usepackage{amssymb, bm}
\usepackage{graphicx}

\begin{document}

\title{Thermoelectric efficiency of topological insulators in a magnetic field}

\author{O.~A.~Tretiakov}
\author{Ar.~Abanov}
\author{Jairo Sinova}

\affiliation{
            Department of Physics \& Astronomy,
	    Texas A\&M University,
            College Station, Texas 77843-4242, USA
}

\date{October 27, 2011}

\begin{abstract}
We study the thermoelectric properties of three-dimensional
topological insulators in magnetic fields with many holes (or pores)
in the bulk. We find that at high density of these holes in the
transport direction the thermoelectric figure of merit, $ZT$, can be
large due to the contribution of the topologically protected
conducting surfaces and the suppressed phonon thermal conductivity.
By applying an external magnetic field a subgap can be induced in the
surface states spectrum. We show that the thermoelectric efficiency
can be controlled by this tunable subgap leading to the values of $ZT$
much greater than $1$. Such high values of $ZT$ for reasonable system
parameters and its tunability by magnetic field make this system a
strong candidate for applications in heat management of nanodevices,
especially at low temperatures.
\end{abstract}

\maketitle

\textit{Introduction.}  Topological insulators (TIs) recently
attracted a lot of attention as potential candidates for spintronics
applications. \cite{HasanRMP10} This interest is based on the
remarkable properties of TIs, \cite{TI_physics_today, Fu07, Hsieh2008,
  Chen2009} namely, the fact that their surfaces (for
three-dimensional TIs) or edges (for two-dimensional TIs) possess
topologically protected conducting states. In this paper we argue that
these unique properties of TIs can also be employed to make very
efficient thermoelectrics. \cite{Murakami2010, Ghaemi10,
  TretiakovAPL10, TretiakovAPL11}

\begin{figure}
\includegraphics[width=0.95 \columnwidth]{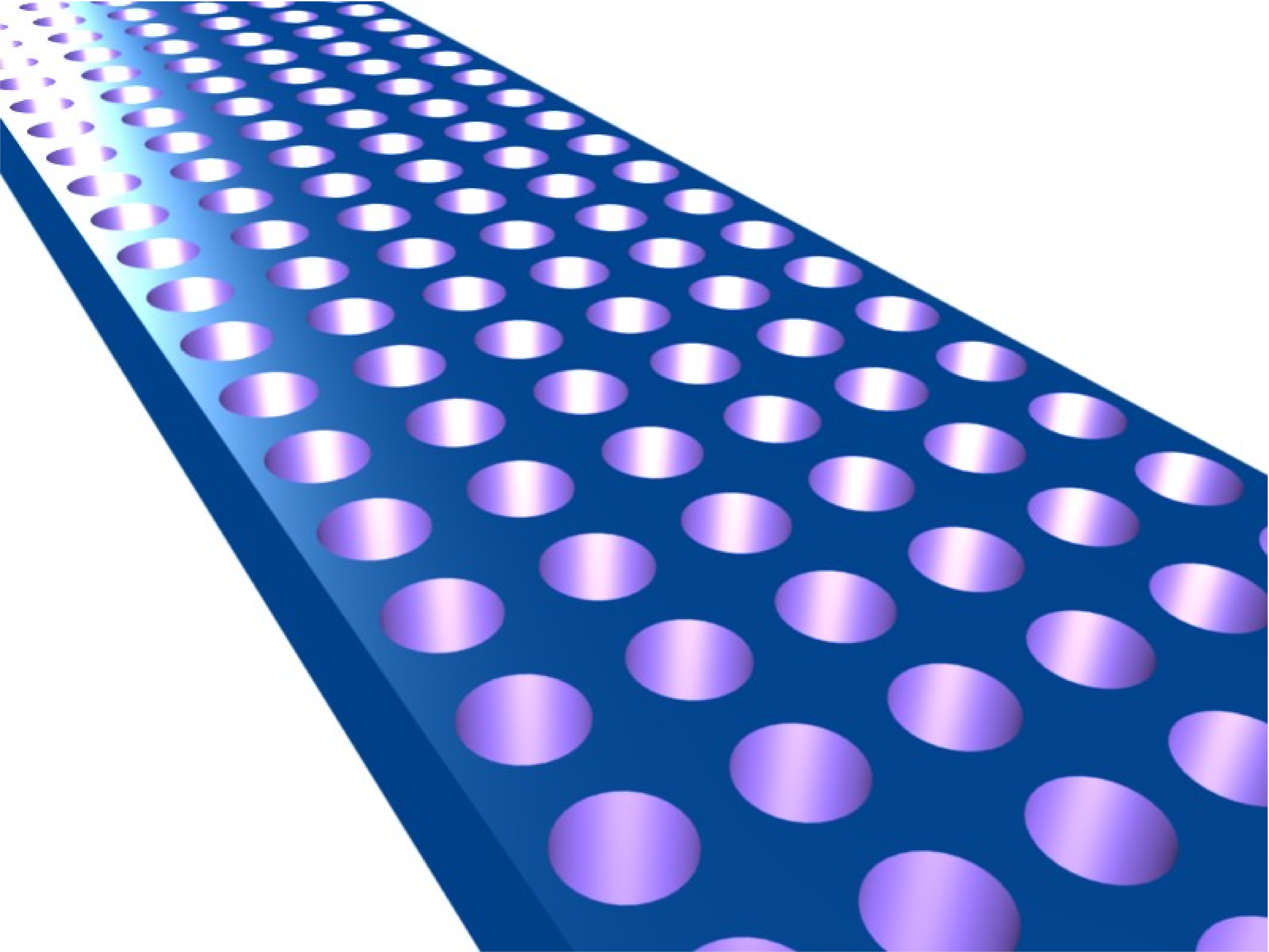} 
\caption{(Color online) A sample of the topological insulator
  with many holes in the direction of transport.}
\label{fig:holey} 
\end{figure}

Good thermoelectric materials \cite{Snyder2008, Tritt99,
  Dresselhaus93, Ho-KiLyeo04, mukerjee07, Markussen09, DiVentra11,
  Shahil10, venkatasubramanian01, zhang:062107, Teweldebrhan10,
  TeweldebrhanAPL10} are characterized by a low thermal conductivity
and high thermopower and electric conductivity. The large thermopower
requires steep dependence of the electronic density of states on
energy, which can be achieved by having the chemical potential close
to the bottom of a band.  The relatively high conductivity demands the
gap to be low. These combined requirements point to semiconductors
with heavy elements as the best candidates for the thermoelectric
materials. Some of these semiconductors were recently rediscovered as
TIs. We show that by making holes in these TI materials (see
Fig.~\ref{fig:holey}) and inducing subgap in their surface states by
means of applied magnetic fields or proximity of ferromagnets one can
significantly enhance their thermoelectric efficiency. The high
density of holes in the direction of transport has two positive
effects: 1) to trap phonons and thus reduce the thermal conductivity;
2) to increase the surface to bulk ratio and therefore effectively
enhance the electric conductivity of the sample. The effect of the
applied magnetic field is to increase the thermopower.

\textit{Thermoelectric figure of merit.} The thermoelectric efficiency
is defined as a ratio of an electric power generated to the total heat
flux through the sample. Therefore, it is characterized by the
dimensionless number
\begin{equation}
ZT = \frac{\sigma S^2}{\kappa_e +\kappa_{ph}} T,
\label{eq:ZT}
\end{equation}
where $\sigma$ is electric conductivity, $S$ is Seebeck coefficient
(or thermopower), $T$ is temperature, and $\kappa$ is thermal
conductivity. The latter, in general, has contributions both from
electrons ($\kappa_e$) and from phonons ($\kappa_{ph}$).

\begin{figure}
\includegraphics[width=0.95 \columnwidth]{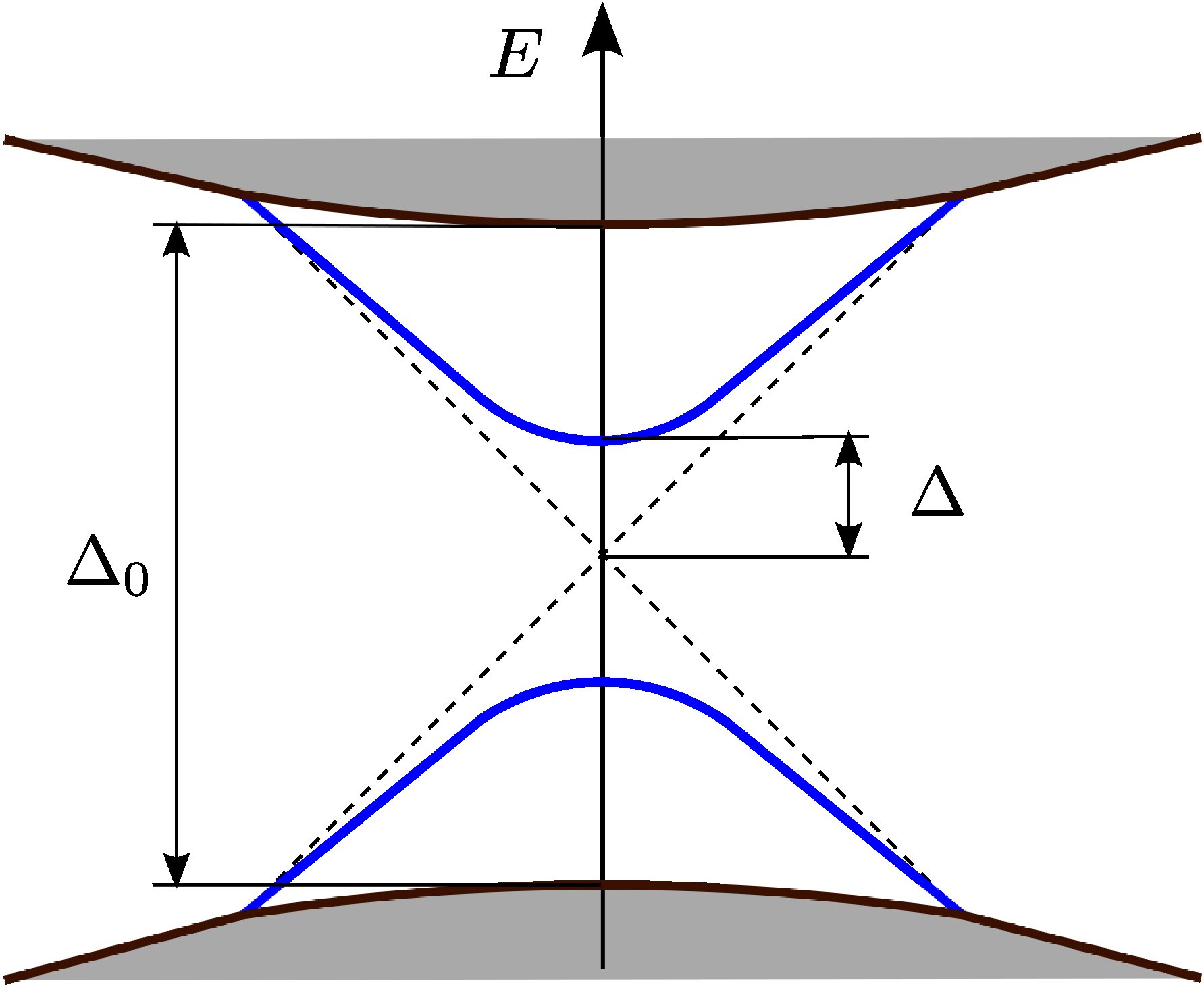} 
\caption{(Color online) A sketch of TI's band structure with the
  subgap $\Delta$ in the surface band. The bulk gap $\Delta_0$ is
  considered to be much larger than $\Delta$.}
\label{fig:subgap} 
\end{figure}

In the absence of magnetic field or magnetic impurities, the TIs have
ungaped surface states. These propagating states are confined to the
close proximity of the surfaces. Their existence is protected by the
topology of the bulk band structure.  These states have cone-like 2D
Dirac spectrum, $E=\pm v\hbar |k|$, where $v$ is the constant Dirac
electron velocity. Application of a magnetic field, doping with
magnetic impurities, or hybridization of these states due to close
proximity of two surfaces can induced a Dirac subgap $\Delta$,
\cite{Analytis2010, Linder09, LuPRB10} which in the case of magnetic
field $B$ is $\Delta\propto B$. Then the surface spectrum takes the
form $E=\pm \sqrt{v^2\hbar^2 k^2+\Delta^2}$, see
Fig.~\ref{fig:subgap}. Generally this subgap is much smaller than the
bulk gap, $\Delta\ll \Delta_0$. For three-dimensional TIs $\Delta_0$
ranges from $0.15$ eV to $0.3$ eV for such as $\rm{Bi}_{2}\rm{Te}_3$
and $\rm{Bi}_{2}\rm{Se}_3$.

\begin{figure}
\includegraphics[width=0.99 \columnwidth]{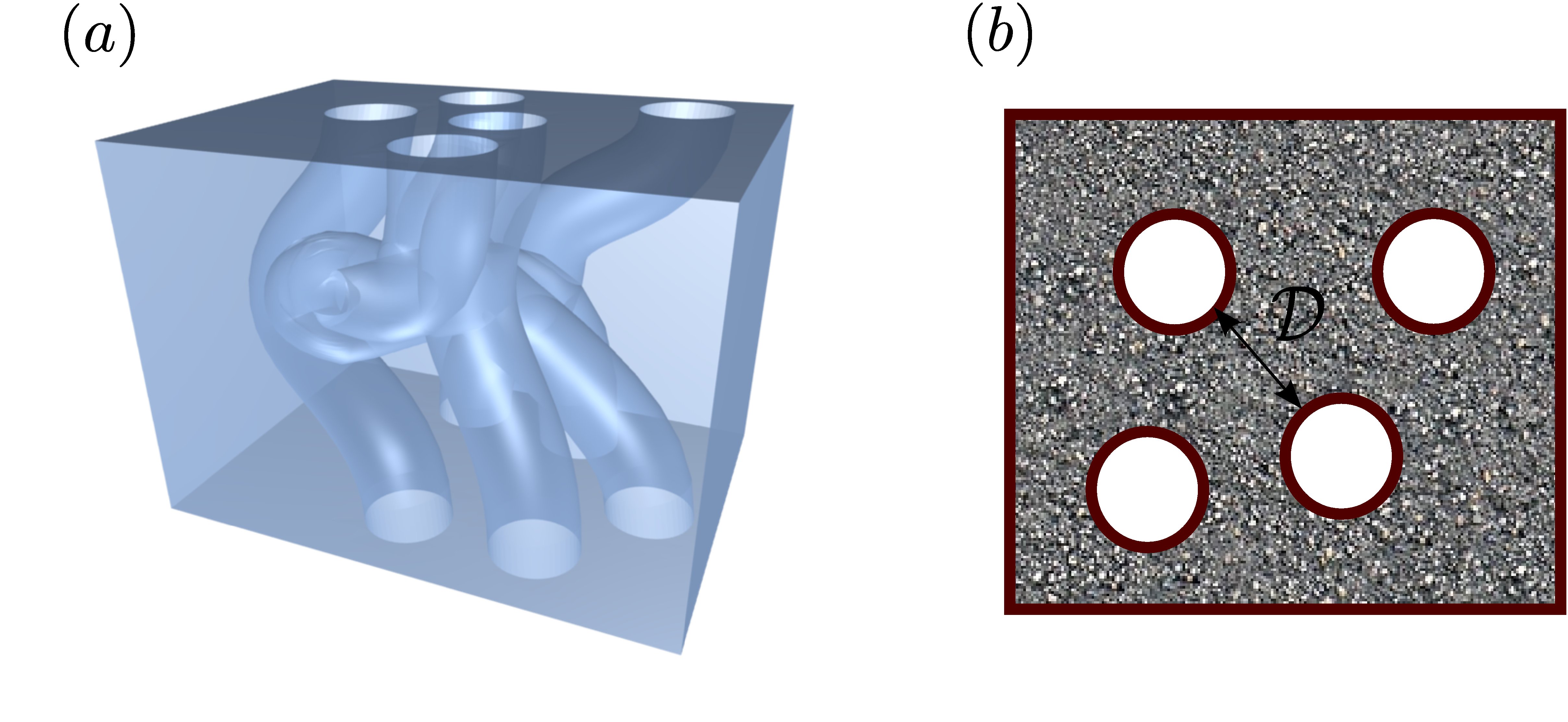} 
\caption{(Color online) (a) A schematic view of a TI sample's part
  with pores propagating in the direction of transport. (b) Top view
  of the TI sample with the average distance between the holes/pores
  given by ${\cal D}$. }
\label{fig:topview} 
\end{figure}

To study the thermal and electric transport in these materials we
employ the linear response theory. \cite{AshcroftMermin} The electric
($j^e$) and thermal ($j^q$) currents are given by linear combinations
of the temperature and chemical potential gradients: 
\begin{eqnarray}
j^e/e= L_0\nabla \mu + L_1 (\nabla T)/T, \\
j^q= -L_1\nabla \mu - L_2 (\nabla T)/T,
\end{eqnarray}
where $e$ is the electron charge.  Using Onsager relations, one can
find from these equations the electrical conductivity $\sigma= e^2
L_0$, Seebeck coefficient $S=-L_1/(eTL_0)$, and electron thermal
conductivity $\kappa_{e}=(L_0 L_2 -L_1^2)/(TL_0)$. The figure of
merit, $ZT$, can then be represented in terms of these linear
coefficients as
\begin{equation}
ZT = \frac{L_{1}^2}{L_{0}(L_{2}+\kappa_{ph}T) -L_{1}^2}.
\label{ZT2}
\end{equation}
In Eq.~(\ref{ZT2}) it is assumed that the transport coefficients have
bulk and surface contributions $L_n = L_{b,n} + L_{s,n}/{\cal D}$,
where ${\cal D}=(A-\sum_n \pi R_n^2)/\sum_n 2\pi R_n$ is the factor
related to surface/bulk ratio (porosity) and has dimension of
length. The holes in the sample does not have to form a periodic
structure and can be placed randomly. The same idea applies to porous
TI samples, see Fig.~\ref{fig:topview} (a). The parameter ${\cal D}$
characterizes the average distance between the pores (holes) of the
average radius $R$, see Fig.~\ref{fig:topview} (b). Here $\kappa_{ph}$
is the phonon contribution to the thermal conductivity in the bulk
(phonon contribution to the thermal conductivity for the surface TI
states is much smaller than that in the bulk).

To estimate the surface contribution of TI to the transport
coefficients $L_n$ we assume the bands to be Dirac-like with a subgap
$\Delta$ and use Boltzmann equation in the relaxation time
approximation,
\begin{equation}
L_{s,n} = - 2 \sum_{i}\int_{-\infty}^{\infty} 
\tau \left(\frac{\partial E_i}{\partial \hbar k}\right)^2  f'(E_i)  
(E_i-\mu)^{n} \frac{d^2 k}{4\pi^2}.
\label{Lsurf}
\end{equation}
Here the sum is over the upper and lower bands, $i=\pm 1$, and
$f' (E)=\partial f/\partial E$ with $f=1/(e^{(E-\mu)/(k_BT)}+1)$ being
the Fermi distribution function. Then taking relaxation time $\tau$ to
be independent of energy, we find
\begin{widetext}
\begin{equation}
L_{s,n} = \frac{\tau (k_B T)^{1+n} }{2h^2} 
\!\!\int^{\infty}_{{\bar \Delta}}\!\!\! dx  
\frac{x^{2}-{\bar \Delta}^2}{x}
\!\left[ \frac{(x-{\bar \mu})^n}{\cosh^2 \frac{x-{\bar \mu}}{2}}
+\frac{(-x-{\bar \mu})^n}{\cosh^2 \frac{x+{\bar \mu}}{2}}\right],
\label{Ls}
\end{equation} 
\end{widetext}
where $h$ is Planck constant, ${\bar \Delta}=\Delta/(k_BT)$, and
${\bar \mu}=\mu/(k_BT)$.

\begin{figure}
\includegraphics[width=0.99 \columnwidth]{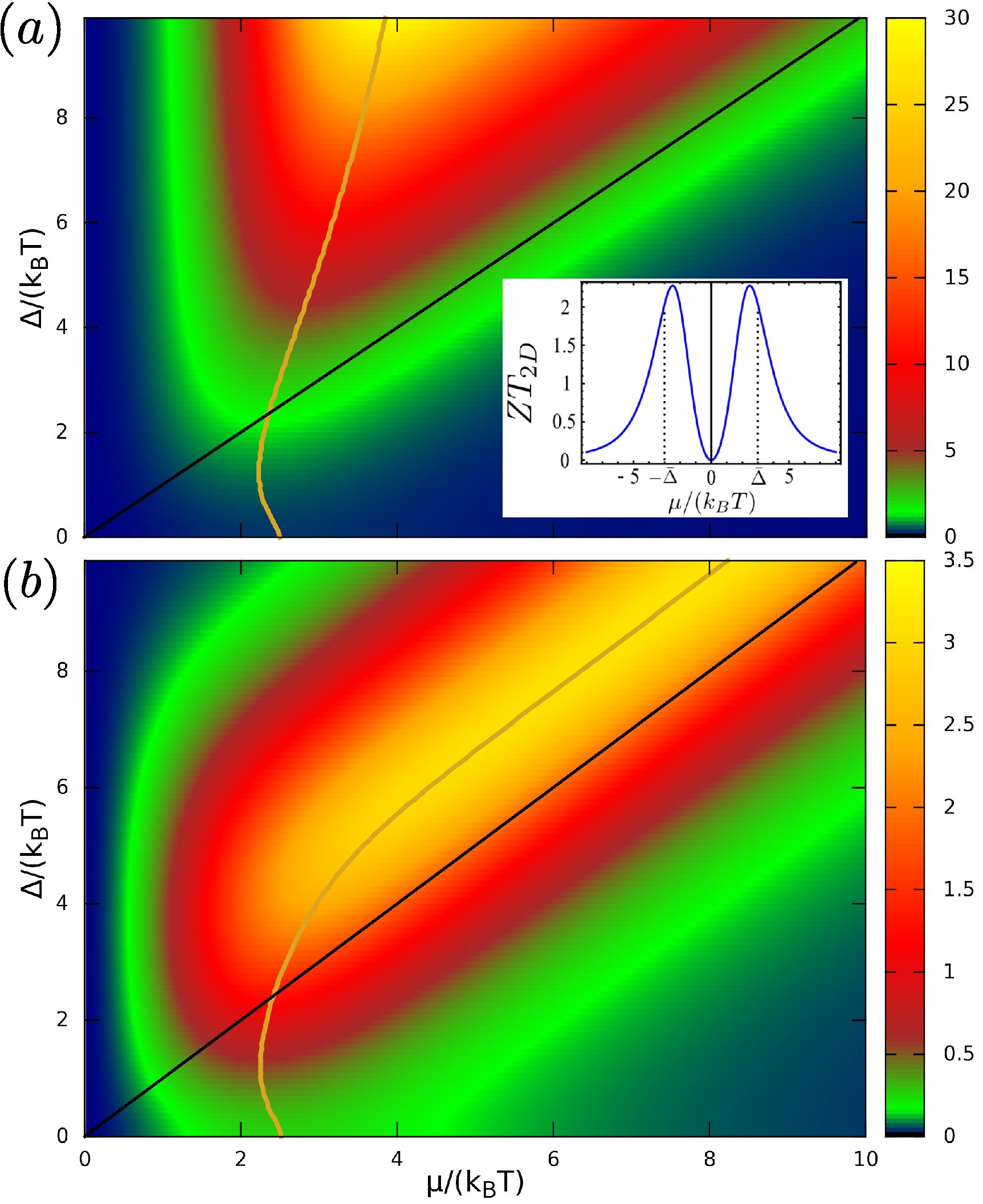} 
\caption{(Color online) Thermoelectric figure of merit, $ZT$, as a
  function of surface subgap $\Delta$ and chemical potential
  $\mu$. Light solid curve shows the maximum of $ZT$. (a) $ZT$ for the
  infinitely high density of holes. The inset shows $ZT(\mu)$ for the
  case of $\Delta/(k_B T) =3$. (b) $ZT$ for a finite density of holes
  characterized by the dimensionless parameter $K_{ph} = 3$. Because
  of the phononic contribution to the thermal conductivity $ZT$ is
  reduced compared to the case (a).}
\label{fig:ZT_all} 
\end{figure}

When the chemical potential $\mu$ is far below the bottom of the
conduction band, $(\Delta_0-\mu)/(k_B T)\gg 1$, the contribution from
the bulk to the transport coefficients is exponentially suppressed,
$L_{b,n}\propto e^{-(\Delta_0 -\mu)/(k_B T)}$, and can be neglected.
The same argument works also for the valence band contribution when
$\mu$ is far from the valence band edge. Thus, the thermoelectric
transport is dominated by the surface states and the only sensible
bulk contribution is to the phonon thermal conductivity
$\kappa_{ph}$. Then the figure of merit becomes
\begin{equation}
ZT=  
\frac{L_{s,1}^2}{L_{s,0}(L_{s,2}+{\cal D}\kappa_{ph}T) -L_{s,1}^2}.
\label{ZT_withKappa}
\end{equation}
At small ${\cal D}$, the contribution to $ZT$ mostly comes from 2D
surface states and in this limit $ZT$ is given by $ZT_{2D}=
L_{s,1}^2/(L_{s,0}L_{s,2} -L_{s,1}^2)$, which is shown as a function
of the chemical potential in the inset of Fig.~\ref{fig:ZT_all} (a)
for $\Delta/(k_B T)=3$ . The color plot of $ZT_{2D}$ as a function of
the induced subgap $\Delta/(k_B T)$ and $\mu/(k_B T)$ is shown in
Fig.~\ref{fig:ZT_all} (a). The maximum $ZT$ achievable by tuning the
chemical potential for a fixed induced gap $\Delta/(k_B T)$ is shown
by a golden line in Fig.~\ref{fig:ZT_all}. The very high values of
$ZT$ in reality are not reachable since any small contribution from
the phonon thermal conductivity reduces $ZT$.

For the holey sample, the phononic contribution to the thermal
conductivity can be characterized by the dimensionless parameter
$K_{ph} = 2\kappa_{ph}{\cal D}h^2/(\tau k_B^3T^2)$. Here the phonon
thermal conductivity can be estimated to be $\kappa_{ph} \approx 1
\rm{Wm}^{-1}\rm{K}^{-1}$ (as for Bi$_2$Te$_3$) \cite{GoyalAPL10} and
the average distance between the holes reaching ${\cal D}\sim
10$~nm. Taking the relaxation time $\tau \approx 10^{-11}$ s, at room
temperatures we estimate $K_{ph} \sim 1$ for these rather conservative
values of parameters. In Fig.~\ref{fig:ZT_all} (b), estimated $ZT$ is
shown as a function of $\Delta$ and $\mu$ for the phononic bulk
contribution characterized by the dimensionless parameter $K_{ph} =
3$. Although reduced considerably from its pure 2D value, $ZT$ remains
substantially larger than any value thus far achieved in these
materials and can be tuned significantly by applied magnetic fields.

\textit{Summary.}  We have studied the thermoelectric properties of
three-dimensional TIs in magnetic fields with high density of holes in
the sample. By applying an external magnetic field a subgap can be
induced in the surface states spectrum. We show that the
thermoelectric efficiency can be controlled by this tunable subgap. We
have find that the thermoelectric figure of merit $ZT$ in these
materials can be much greater than $1$. This is due to the high
contribution of the topologically protected conducting surfaces and
the suppressed phonon thermal conductivity.  High $ZT$ values for
reasonable system parameters and its tunability by magnetic field or
magnetic impurities make this system a strong candidate for
applications in heat management of nanodevices.

We thank M. Bakker, J. Heremans, C. Jaworski, J.~E. Moore, O.~Mryasov,
and D.~Pesin for insightful discussions. This work was supported by
NSF under Grant No. DMR-1105512, 0757992, DMR-0820414,
ONR-N000141110780, SWAN, and by the Welch Foundation (A-1678).

%

\end{document}